\documentclass{article}
\usepackage{hyperref}
\usepackage{graphicx}
\usepackage{fullpage}
\usepackage{verbatim} % Support for the 'comment' environment.
\usepackage[utf8]{inputenc} 
\usepackage{authblk}

\providecommand{\keywords}[1]
{
  \small	
  \textbf{\textit{Keywords---}} #1
}
\newcommand{\etal}{\textit{et al.}}
\title{Digital Product Passport Management with Decentralised Identifiers and Verifiable Credentials - Extended Version}
\author[1,\footnote{Corresponding author. E-mail address: iillan@iti.es}]{Ismael Illán García}
\author[1,2]{Francesc D. Muñoz-Escoí}
\author[1]{Jordi Arjona Aroca}
\author[1]{F. Javier Fernández-Bravo Peñuela}

\affil[1]{\small{Instituto Tecnológico de Informática, Camino de Vera, s/n, 46022, Valencia, Spain}}
\affil[2]{Universitat Politècnica de València, Camino de Vera, s/n, 46022, Valencia, Spain}

\date{}

\begin{document}

\maketitle
%\tableofcontents

\begin{abstract}
Digital product passports (DPP) have been proposed in the European \textit{Ecodesign for Sustainable Products Regulation} (ESPR) as a means to keep and provide product information that facilitates product reusage, reparation, and recycling. Thus, DPPs should provide a positive effect on the environmental impact of future manufactured products, preventing waste and promoting a circular economy (CE) model.

ESPR settles a set of requirements in collecting and administering product-related data. Decentralised identifiers (DID) and verifiable credentials (VC) are two self-sovereign-identity-related elements that may help in that DPP management since they introduce a decentralised administration of identity that may enhance the overall scalability of the resulting system, improving also its reliability. This paper analyses the ESPR requirements and describes how they may be achieved using DIDs and VCs, assessing their performance in some scenarios.
\end{abstract}


\section{Introduction}

Digital product passports (DPPs) provide a verifiable collection of data about products' composition, environmental impact and opportunities for preventing waste \cite{Lang23}. Thus, the DPP should be a key element in the development of a Circular Economy (CE) and this has led the European Commission \cite{EC24} to a regulation on what DPPs shall be in order to implement and use them in a few years.

Although CE initially referred to an economy system based on a three-action strategy, usually known as 3R (i.e., Reduce, Reuse and Recycle), recent proposals (e.g., Potting \etal~\cite{Potti17}) have extended it to a larger number of complementary actions (9R): R0 Refuse (Make products redundant by abandoning their function), R1 Rethink (Redesign to make product use more intensive), R2 Reduce (Consume fewer natural resources while the product is manufactured or used), R3 Reuse (Other consumers may use again a product that is still in good condition), R4 Repair (Repair a defective product instead of discarding it), R5 Refurbish (Restore an old product), R6 Remanufacture (Use parts of discarded products in a new product with the same function), R7 Repurpose (Use parts of discarded products in a new product with a different function), R8 Recycle (Process materials to obtain the same or lower quality) and R9 Recover (Incineration of materials with energy recovery). There, R0 to R2 promote smarter product use and manufacture, R3 to R7 extend the lifespan of products or their parts while R8 and R9 advise on a useful application of materials. The information kept in a DPP assists the different agents that participate in those actions (e.g., suppliers, manufacturers, retailers, customers, repairers, recyclers...) to build an efficient CE ecosystem.

But the implementation of DPPs will not be trivial. There are many facets to consider: (i) their information should be accessible by the customers, to easily choose the most adequate (in regard to environmental footprint) alternative when any product should be bought, (ii) manufacturers should state in the respective DPP the composition of every product they sell, but that information --in some specific manufacturing areas-- may have been kept traditionally secret and may provide some advantages to their competitors, (iii) recyclers may need multiple data from DPPs, but not necessarily those of interest for the customers... In the end, different users of the DPP need different data from it. How can we design, implement and manage a tool like this?

Decentralised identifiers (DIDs) \cite{Sporn22} and verifiable credentials (VCs) \cite{Sporn19,Sporn22b,Sporn23} may provide a good basis to solve those problems. DIDs are a specific kind of identifier designed for supporting self-sovereign identity (SSI) \cite{Allen16}. In a SSI system, each individual may generate and use its own DID(s) when needed, and associate information to them with VCs. The novelty is that such identity-related information is now managed and presented by the identified individual when it considers, instead of being kept and administered by external companies or services, like in previous centralised or federated identity systems. Now, each person or company will have its own DID and complementary data in the form of VCs, deciding at each moment which data is presented to others (through verifiable presentations, or VPs \cite{Sporn19,Sporn22b,Sporn23}) when those others demand that information in a given interaction. Additionally, those third parties will not be able to present later that data to other parties, since each action that involves identity-related information uses a VP that should be signed by the sender and it may be easily rejected and disregarded when the VP signer is not the subject of the VP. Therefore, with DIDs, VCs and VPs, persons and companies may easily control which of their own information is presented to others and when and how this is done.

\begin{figure}[h]
\centerline{\includegraphics[width=0.8\textwidth]{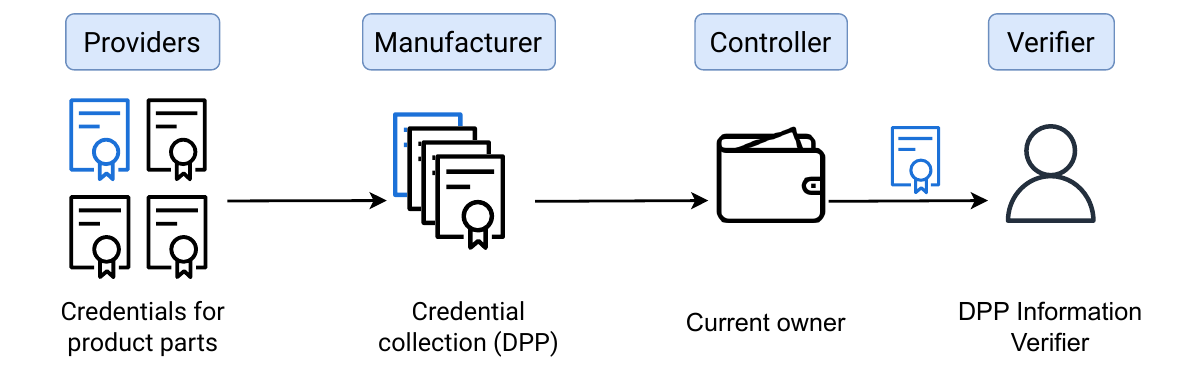}}
\caption{Composition of DPP information from its components and the selective disclosure of specific credentials from the owner to the verifier.}
\label{fig:dpp-vc}
\end{figure}

Additionally, DIDs may also refer to inanimate entities, like manufactured products, and in that case their associated VCs and VPs will be handled by a given controller, specified at DID creation time. In the general case, the controller for a given product may be its manufacturer or its current owner. Thus, when the information of a product, that has an identifying DID, is stated in multiple VCs, each one devoted to a specific part, the DID controller may easily choose which VC is taken for preparing the corresponding VP and deliver that VP to a given requesting party. Following that strategy, a DPP may be implemented as a collection integrated by a DID and several VCs, with their respective VPs that present different parts of the DPP (i.e., a concrete subset of the existing VCs) to each requesting agent (e.g., retailers, users, repairers, recyclers...) depending on the involved product lifecycle stage, as Figure \ref{fig:dpp-vc} shows.

Therefore, we consider that the main contributions of this article are as follows. First, the analysis of  ESPR~\cite{EC24}, since being an extensive legislative document, it requires considerable effort to read and process all its sections. From this analysis, the next contribution arises: see the requirements imposed by this regulation for the implementation of a digital product passport. Finally, it is demonstrated how this passport could be implemented based on the decentralised identity paradigm, which we believe aligns quite accurately with the previously identified requirements.

To this end, the paper is structured as follows. Section \ref{sec:sota} discusses the current state of the art in this research area. Section \ref{sec:reqs} explains the requirements for identity-related management derived from ESPR. Section~\ref{sec:analysis} analyses in detail the problems that need to be solved to implement a DPP system and provides a first draft of a DID-based design that solves each of them. Section \ref{sec:proposal} describes how those mechanisms should be implemented. Section \ref{sec:discussion} provides a brief discussion on the quality of our proposals and an analysis of their costs. Section \ref{sec:use_cases} describes several use case scenarios and, finally, Section \ref{sec:conclusion} provides the conclusions.

\section{State of the Art} \label{sec:sota}

%To analyse whether DIDs should or could be used to identify products in their DPPs, we will structure our presentation as follows: Section \ref{normative} studies the implications of the European regulations we described previously and Section \ref{solutions} explains the results generated in other recent academic publications.

Three different dimensions should be considered in a review of the current state of the art on the usage of DIDs in the future implementation of DPPs: a brief summary on self-sovereign identity (SSI) management (Section \ref{ssi}) that introduces the basic terminology to be used in subsequent sections of this paper, the existing EC regulations on DPPs (Section \ref{normative}) and the preliminary results generated either by pilot projects or other academic research in this field (Section \ref{solutions}).

% Should we include a first subsection describing a DID-VC-VP ecosystem? 

\subsection{Self-Sovereign Identity} \label{ssi}

SSI~\cite{Allen16} refers to a decentralised identity management where each individual may control and administer what information others have regarding its identity, as opposed to centralised or federated identity managements, where that control belongs to a third party. The World Wide Web Consortium has published several recommendations defining what should be understood as a DID~\cite{Sporn22} and how those identity-related data may be transferred among agents~\cite{Sporn22b}, using VCs and VPs to this end.

\begin{figure}[h]
\centerline{\includegraphics[width=0.62\textwidth]{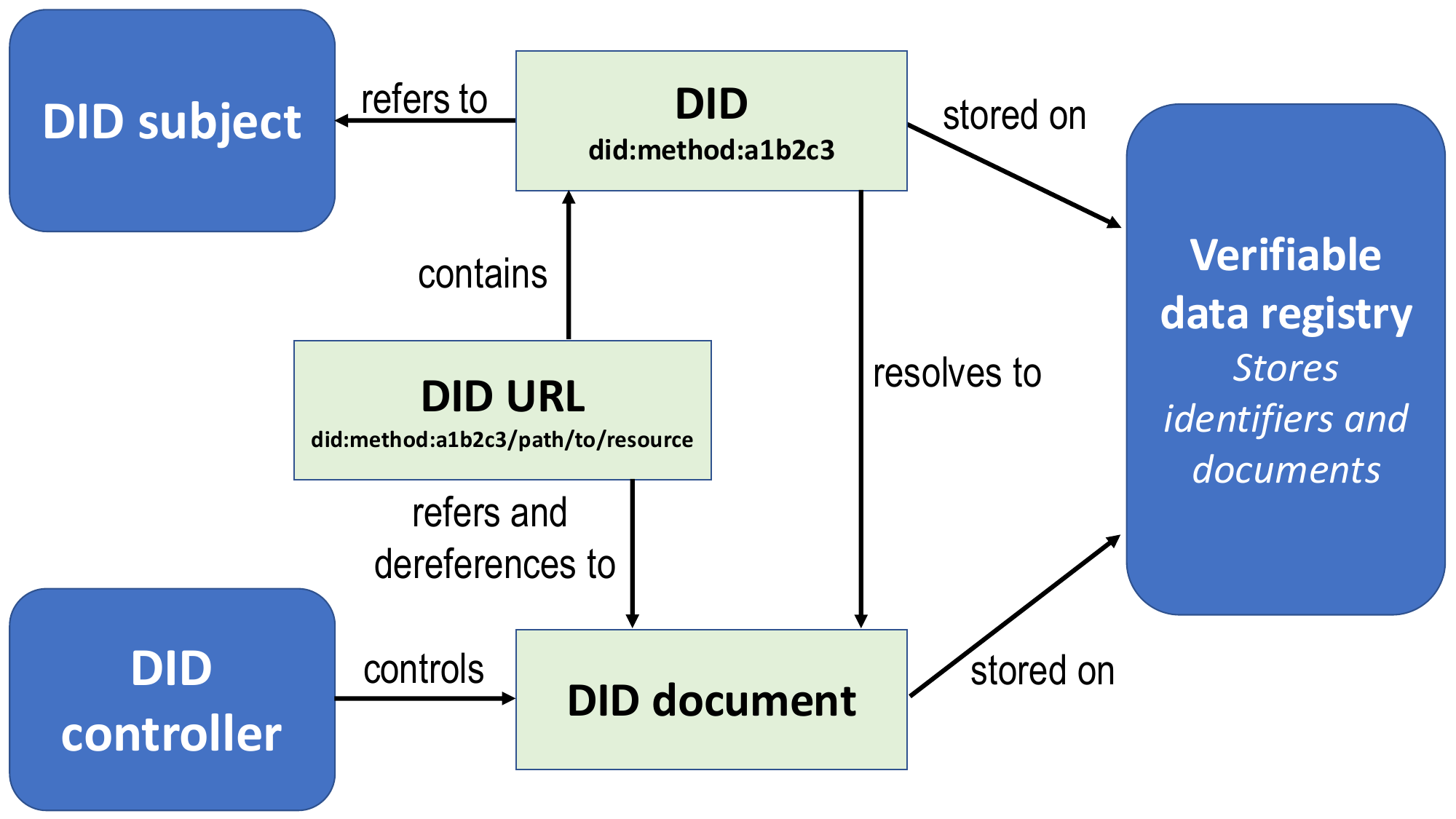}}
\caption{Basic components of a DID architecture~\cite{Sporn22}.}
\label{fig:did}
\end{figure}

The main components in a SSI system based on DIDs are depicted in Figure \ref{fig:did}. The central element is the DID. A DID is a URI~\cite{Berne05} composed of three parts separated by colons: (a) the scheme (invariable, it should be ``\textbf{did}''), (b) the method name, and (c) a method-specific identifier. A DID identifies a given entity: the DID subject. When a DID is resolved, we obtain a DID document that stores some information on the DID subject like the cryptographic public keys needed in different verification methods. That information kept in the DID document may be updated by the DID controller. Usually, the DID subject and the DID controller are the same entity, but they may be different when the DID refers to, e.g., inanimate entities, like industrial products.

\begin{figure}[h]
\centerline{\includegraphics[width=0.62
\textwidth]{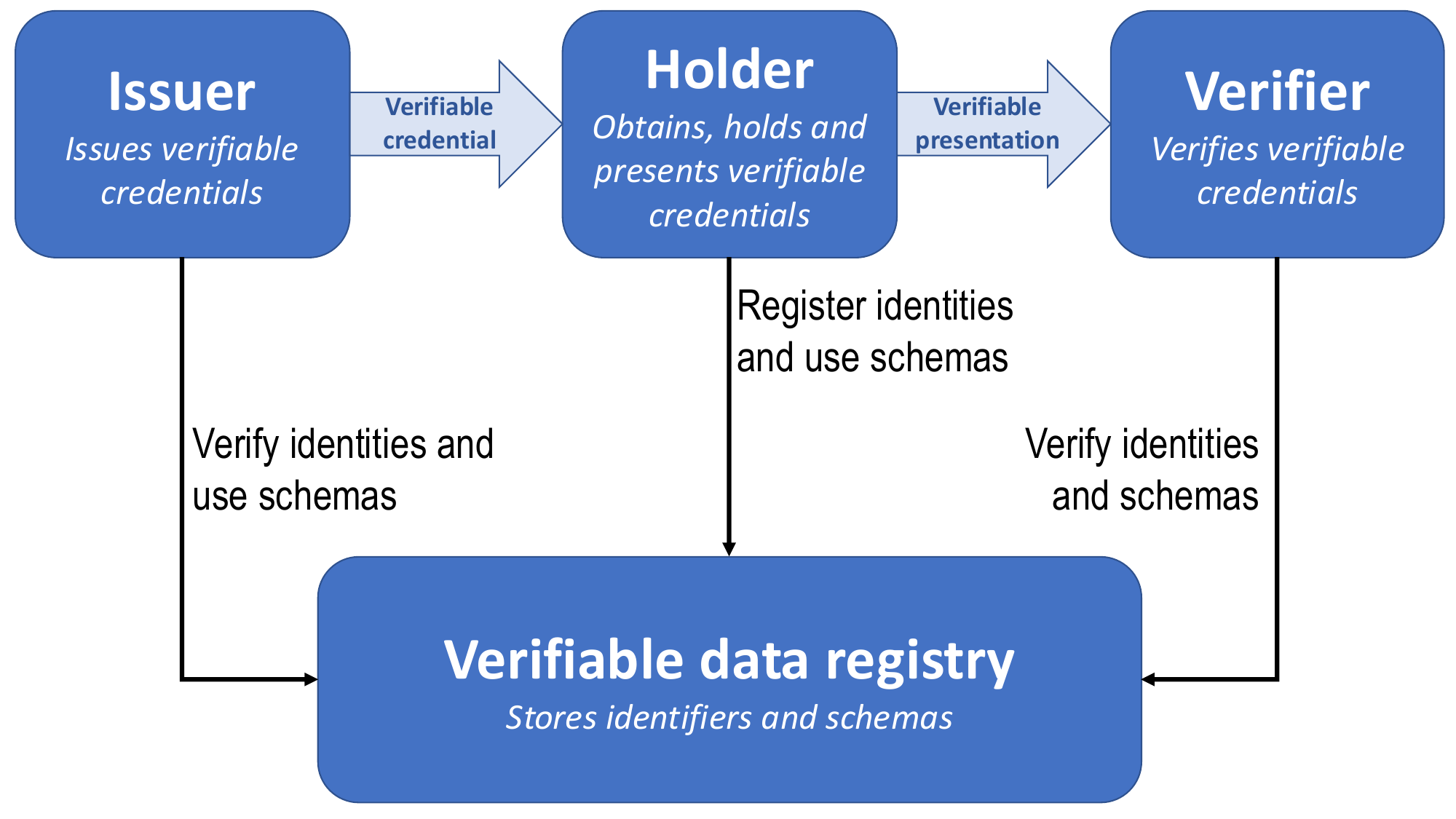}}
\caption{Roles and data flows in VC-based interactions~\cite{Sporn22b}.}
\label{fig:vc}
\end{figure}

In order to refer to a specific field of the DID document, we may use a DID URL that extends the DID with a pathname that refers to that intended field. DID documents are stored on verifiable data registries (VDR) that keep all that information persistently (e.g., using distributed ledger technology (DLT), i.e., a blockchain~\cite{Fern21}).

\begin{figure}[h]
\centerline{\includegraphics[width=0.8\textwidth]{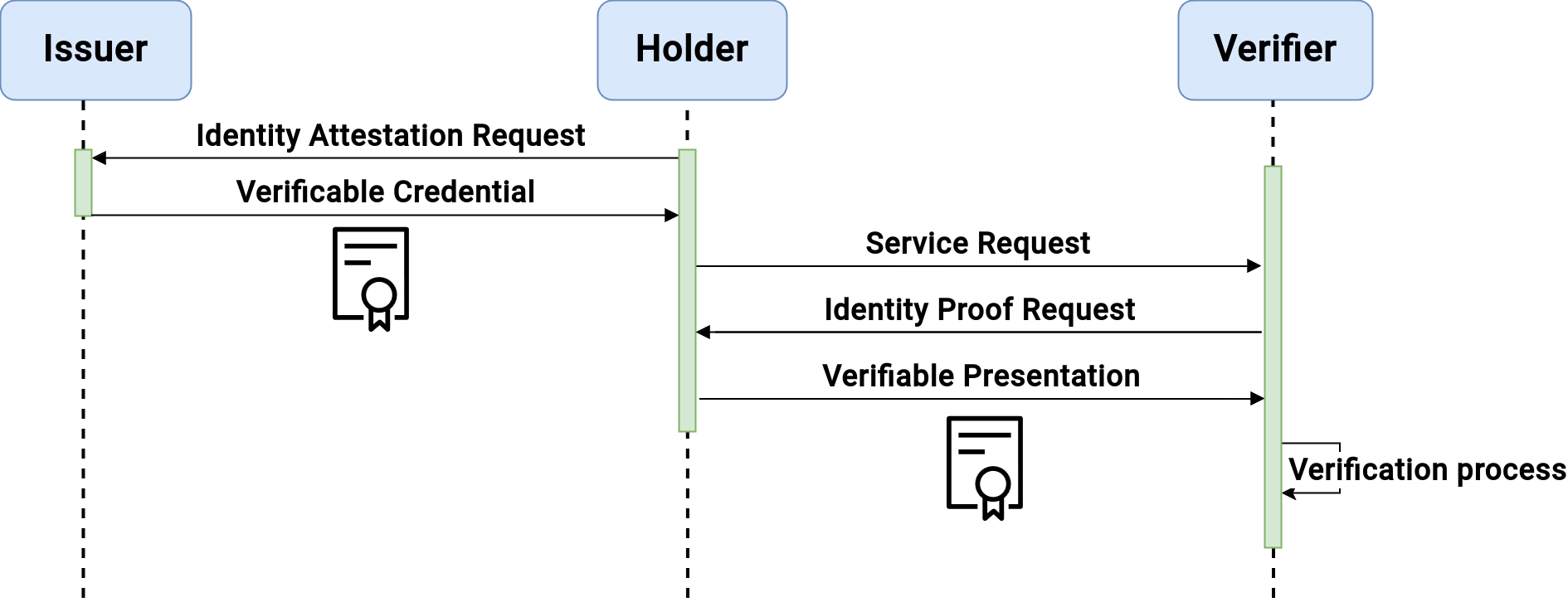}}
\caption{Sequence diagram in VC-based interactions.}
\label{fig:vc-sequence}
\end{figure}

The information related to a given identity may be accredited or requested by other agents, as it is depicted in Figures \ref{fig:vc} and \ref{fig:vc-sequence}. In that case, such an information is stated in a VC and signed by an issuer A, who generates that VC and delivers it, using a specific protocol (e.g., \cite{Khate21,Lodde24}), to the VC holder B that keeps it in a digital wallet. In the regular case, B will be the DID subject entity presented in Figure \ref{fig:did}. Later on, B may be required to prove some identity-related claim in order to get some service from C. To this end, C may demand B some proof about that claim and B will present a VC stating that such claim is true. So, B takes the VC and signs it, generating thus a VP that is delivered to C using another specific protocol (e.g., \cite{Khate21b,Terbu24}). Now, C behaves as a verifier and recovers B's public key from the VDR in order to check the VP signature. If that check is passed, then C assesses whether the original VC signature corresponds to A, reading A's public key from the VDR. If all those verifications are successful, C provides the services that B requested.

\subsection{Regulations on Digital Product Passports} \label{normative}

The European Commission (EC) has been a pioneer in requiring the use of DPPs. To this end, ESPR \cite{EC22} was drafted in March 2022, supported by several funding calls (e.g., \cite{EC23}) for pilot implementations in various fields. In December 2023, the European Parliament and the Council reached an agreement on that regulation~\cite{EC23b}, thus passing the next required stage to reach its final adoption, that occurred on June 2024 \cite{EC24}. Once adopted, EC should still choose and regularly update and expand the list of products to be affected by that law, publishing subsequent delegated acts following a working plan that shall cover a period of at least three years and be regularly updated (as stated in its Article 18.3~\cite{EC24}). %Note that the main goal of this ESPR is to enhance and extend the regulations of the previous (and still active when this paper has been written) European Ecodesign Directive (EED)\cite{EC09}, and the latter already planned some acts for subsequent product groups (e.g., computers and computer servers at the end of 2025 \cite{Nicke24}). 

ESPR is a large document that consists of fourteen chapters, subdivided in eighty articles, and complemented with eight annexes. Its first article may be summarised as follows~\cite{EC22}: \textit{``Article 1 lays down the subject matter of this Regulation, namely a framework for setting ecodesign requirements, creating a digital product passport, and prohibiting the destruction of unsold consumer products. It lays down the product aspects to which the eco-design requirements relate, such as durability and reliability, reusability, upgradability, repairability, and possibility of maintenance and refurbishment, presence of substances of concern, energy and resource efficiency, recycled content. It further sets the scope of the Regulation...''} This explicitly shows that the first goal of the ESPR is to drive the European economy to a circular model.

Chapter III, consisting of Articles 9 to 15 is directly related to DPPs since it states their definition, requirements, technical design and operation, what  unique identifiers are, the existance and management of a product passport registry, the existance and management of a web portal for data in DPPs, and customs controls related to the DPP, respectively. Finally, Annex III outlines which data elements shall or may compose a DPP.

There are twelve data elements in that annex, and that list provides an orientation on the information to be included in a DPP, but the exact contents may depend on the specific product group and on the future delegation acts for each industrial area. The third element refers to product identification and it requires \textit{``the Global Trade Identification Number as provided for in ... standard ISO/IEC 15459-6 \cite{ISO14e} or equivalent of products or their parts.''} However, that statement does not prevent manufacturers from complementing the Global Trade Identification Number (also known as Global Trade Item Number~\cite{GS1_24}, GTIN) with a product DID. Note also that the second data element in Annex III refers to another unique product identifier as indicated in the corresponding applicable delegated act. Thus, those subsequent acts may extend the type of identifiers accepted for each product group. Therefore, an analysis about the convenience of DIDs in a DPP still makes sense.

Multiple ESPR sections define what should be and do a DPP. From them, these may be of specific interest to our paper goals:

\begin{itemize}
\item The three last points in Article 5.11 state that there should not be any disproportionate negative impact on the competitiveness, nor proprietary technology imposition, nor disproportionate administrative burden on economic actors (and among them, mainly, the manufacturers) because of the introduction and management of DPPs.

\item Article 9.2 states that there should be specific subsequent delegation acts for each product group (i.e., manufacturing area) that  consider concrete subsets of DPP management. To this end, multiple points are presented in that article. Let us focus on the following:
\begin{itemize}
\item Point (d) analyses whether the product passport should correspond to the model, batch or item level. This means that those decisions do not correspond to the manufacturer, but they will be explicitly taken in those subsequent laws.
\item Point (f) mentions \textit{``the actors that are to have access to data in the digital product passport and to what data they are to have access,''} depending on their role: customers, end-users, manufacturers, importers, dealers, recyclers...
\item Point (g) specifies \textit{``the actors that are to create a digital product passport or update the data in a digital product passport and what data they may introduce or update.''} Thus, it will be ruled and controlled who may create or update a DPP.
\item Point (i) refers to the period for which the DPP shall remain available.
\end{itemize}

The existance of subsequent delegation acts for specifying some concrete DPP management details for each product type has reduced the focus of academical research on DPP contents, since the product attributes and characteristics to be handled in those DPPs will depend on those future delegation acts. In spite of this, some relevant papers have been published in this regard \cite{Berge22,Jense23}.

\item Article 11.(g) requires that the DPP data authentication, reliability and integrity shall be ensured.

\item Article 11.(h) demands that \textit{``digital product passports shall be designed and operated so that a high level of security and privacy is ensured and fraud  is avoided.''}
\end{itemize}

% We should still discuss the adequacy of DIDs and VCs in order to comply with these requirements, but that discussion could be presented at the "Analysis" section.

% Other normative, besides the European one, may exist on circular economy and perhaps it may suggest some interest in DPPs. We should look for it and summarise the results here.

Section \ref{sec:reqs} rewrites those conditions into requirements and Section \ref{sec:analysis} analyses in depth how DIDs and VCs may be able to support and implement all those requirements, discussing their advantages when those mechanisms are compared with other alternatives. 

\subsection{Solutions} \label{solutions}

% Perhaps we should start with some proposals that are not based on DIDs, to simply discuss that DPPs may be implemented and have been implemented in pilot projects, but their implementation is not trivial at all. Later on, we should revise the Guth and Navarro works, without so many criticisms. In the end, if there are drawbacks in their proposals, they have been mainly caused by the existing uncertainty. In general, their proposals make sense, and the concrete details to be respected by DPPs will be stated in future specific delegated acts.

% It does not make sense to cite Voulg24. It does not provide any serious proposal.
Although European CE regulation proposals raised almost immediately a high interest on DPPs \cite{Adiso21,Afrid23,Berge22,Duran22,Jense23,King23,Lang23,Psaro24,Strat23,Walde21}, there have been few proposals that discuss the design or implementation of DPPs through DIDs. This is because the very concept of a digital passport is recent, and there are no adopted regulations yet that govern its structure and the necessary technologies for its development and use.

One of the initial proposals has been that of Dr. Susanne Guth-Orlowski \cite{Guth21, Guth22, Guth23}. She states that these developments are feasible and offer a significant number of advantages. Her proposal does not focus exclusively on DIDs and VCs; other technologies could be used to this end, but she does provide a primary example based on decentralized identifiers and verifiable credentials for batteries. As described in \cite{Guth21}, all the companies involved in the lifecycle of a specific product (component manufacturer/supplier, product manufacturer, dealer, repairer, recycling centre, waste disposal centre) may have their own DID for identification. Both suppliers and manufacturers use VCs to declare the materials that make up each product component. The manufacturer aggregates these credentials to create the first version of the passport for a specific reference. This reference has its own DID, and the credentials may be issued by the manufacturer using the DID of that reference as their subject. On this DID, external auditors emit certifications to be included in the digital passport. The author suggests using the same identifier for product batches and, if the composition of the batches generated over an extended period does not change, reusing the same passport for all of them. This proposal pays particular attention to the initial composition of a product and is less focused on potential changes that may be applied during the useful life of a specific product unit (since that seldom happens in the battery market) before it reaches the recycling stage. The same approach is applied by Berg \etal \cite{Berg22} on batches of plastic products.

% fmunyoz: Both Berg22 and Guth21 mention the future existence of the DPP registry, whose need and functionality is outlined in the ESPR proposal. We should also explain that element and relate it to the public availability of DPPs.
A problem that arises in \cite{Berg22,Guth21} is the mention that credentials will be kept in the DLT (i.e., in the verifiable data registry, or VDR, of the corresponding DID-based system) and can be queried using the DID resolution mechanism. However, in the World Wide Web Consortium (W3C) standard \cite{Sporn22}, such a resolution operation obtains the DID document, which contains metadata (mainly, the public key of that identified entity), but not the credentials. Nevertheless, that DID document may maintain one or more service endpoints to access the services associated with that identified entity. In \cite{Guth21}, it is assumed that one of the services would be the link to the storage location of the verifiable credentials. According to W3C \cite{Sporn22b}, credentials should be kept in the digital wallets of DID owners, and their issuer would generate them exclusively for that owner. Since products are not active agents able to initiate interactions, what is recommended in the standard must be adapted to fit the common scenario for managing passports. To make everything fit, subsequent articles \cite{Guth22, Guth23} mention digital twins, so it can be assumed that those twins behave as the DID controller when managing those credentials that make up the DPP. In any case, the manufacturer could actually manage some wallet \cite{Guth22} (either its own or those associated with digital twins) that maintains all these credentials. Therefore, anyone who needs to access the DPP (e.g., a recycling centre) would interact with the wallet maintained by the manufacturer.

In \cite{Guth22}, the main use of the DPP is still for the user to query the initial composition of a product, but at the end Dr. Guth-Orlowski states that further work is needed about how to add usage information to a product. In this design the end user is not the controller of the product DID, so she can not issue a VC about certain product and add it to the DPP. Dr. Guth-Orlowski proposes that the end user receives a delegation key with limited rights to sign VCs to the product DID.
\begin{comment}
% TO DO: This discussion should be moved to the Analysis section!!!
With this proposal, a series of new questions arise:

\begin{itemize}
	\item How would be controlled the creation of the delegation keys? The user of a product may receive a key when acquiring the product but in most cases he will not emit any verifiable credential about the product. Instead an authorized agent, like a mechanic, may issue a credential about the usage state of a product.  
	\item In the example above, the mechanic would sign the verifiable credential about the product with their own private key. Where will this credential be stored? Should the manufacturer maintain the state of the digital passport as well as its updates during the product life cycle? Or should the current owner be the one in charge of storing the new verifiable credentials in their own wallet? The latter supposes that to obtain all the information about a certain product, a user would have to request the digital passport from the manufacturer and the credentials obtained during the lifecycle of the product from the owner. Furthermore, it would be necessary to implement mechanisms to ensure that the user is sharing all the credentials they possess, as they could conceal some information that might hinder its sale.
\end{itemize}
\end{comment}

Moreover, the protocols to be used for interaction between wallets in a DID-based system are not specified in the W3C recommendations. Therefore, this author mentions \cite{Guth23} some alternatives (DIF DIDComm Messaging V2, DIF Presentation Exchange V2.0, OpenID Connect, OID4VC, OID4VP...). The fact that W3C has specified a metasystem means that there are many viable approaches to managing a particular type of interaction. This should facilitate interoperability with existing systems already in production, but it complicates the development of hypothetical generic wallets due to the high number of protocols to consider and the diversity of configurations that each one supports.

A second proposal has been developed by Leandro Navarro et al. \cite{Navar22}. In this case, it does not analyse the management of DPPs in a general sense but focuses on a specific scope: digital devices. These devices often consist of multiple digital components, each with its own reference and composition. Therefore, in this case, a complex product must be managed, created by assembling a large number of parts, each with its own information about the materials used in its manufacture. Additionally, during their lifespan, these devices may malfunction and require the replacement of some of these components to complete their repair.

Similar to Guth-Orlowski, Navarro advocates the use of DIDs to identify these devices. However, there are significant differences compared to Guth-Orlowski's proposal. The first difference is that the DPP will not be a set of verifiable credentials but another ''identifiable entity'' and, therefore, will have its own DID, derived from the DID of the digital device it provides information about. This duplicity also affects the DIDs to be managed, which will use two different methods in their respective URIs, resulting in the following DIDs:

\begin{itemize}
    \item did:eCHID:\textit{CHID} for the products.
    \item did:eDPP:\textit{CHID}:\textit{PHID} for the passports.
\end{itemize}

Here, ``chid'' refers to the ``chassis'' identifier, which represents the original configuration of a digital device. Typically, this identifier would consist, in a specific sequence, of the manufacturer's identifier, the product's reference number within that manufacturer's catalog, and the serial number. On the other hand, ``phid'' (from ``product hardware identifier'') will be a number generated by a certain benchmarking application that considers each of the pieces that make up that device. When a component is replaced during a repair, it will result in a modification of the ``phid'' that identifies the device.

Therefore, when the device is created, its manufacturer will generate its DID, as well as the initial version of its DPP. The DID document for its DPP will contain all the relevant information about the materials that make up that device. Perhaps this initial description includes the information present in the DPPs associated with its components, if they exist. Every time, during its lifespan, the device undergoes a change in any of its elements, a new DPP will be generated (as there will be a change in the final component of its DID: the ``phid''), along with its associated document, and it will be stored in the VDR.

As we can see, this second proposal does not manage VCs, and it is not documented that traditionally associated public keys of a DID are used. Instead, the relevant information about materials used will be directly stored in the DID document, but not for the product's DID, but for the DPP's DID. By using this architecture, there is no need for the use of digital twins to manage credentials or private keys, as DIDs are not used for authentication or credential management but solely to access persistent information present in DID documents.

\section{Requirements} \label{sec:reqs}
Section \ref{normative} has enumerated several articles in the ESPR that are somehow related to identity or to access permissions and that could determine several requirements in the management of product identities or entity accesses. Let us translate now those articles into specific requirements:

\begin{enumerate}
\item[R-01] \textit{\underline{Product identifiers} should comply with the standard ISO/IEC 15459-6}. Those identifiers are also known as global trade item numbers (GTINs). This requirement is derived from Annex III of ESPR. Therefore, DIDs, if any, should complement GTINs, and be another data element in the DPP, instead of replacing them. Anyway, VCs and VPs may be associated to other identifiers (e.g., GTINs), when no DIDs are used.

\item[R-02] \textit{A \underline{DPP} generic implementation may support model, batch or item \underline{granularity}, but for a specific product type such granularity will be already determined by its specific delegation act}. Article 9.2.(d) of ESPR states that rule. So, DPP granularity cannot change along time in a given type of product. 

\item[R-03] \textit{\underline{DPPs should be adequately stored and maintained} along the product life cycle}. Thus, at least one agent should carry out the involved tasks, i.e., one (or, perhaps, a given subset) of the actors in a DPP system should be responsible of keeping the DPP data along the life cycle of the products. Besides, that responsibility may dynamically change from one actor to the next when a product makes a transition to a different stage of its life cycle (e.g., a person sells its car to a garage, where it will be carefully revised, repaired if needed, and put again for sale; in a scenario like this, DPP maintenance responsibilities might be temporarily assigned to the garage). This is not a direct requirement extracted from the ESPR articles, but it conditions how to implement some solutions to other subsequent requirements (e.g., R-04, R-05, R-06 and R-09).

\item[R-04] \textit{\underline{DPP data shall remain available} for a specific time interval}. The length of that interval depends on the product group being considered and will be determined in the corresponding delegated acts, but it should be, at least, long enough to preserve the DPP information until the complete life cycle of all the products referred in that DPP has been exhausted, thus making possible that the corresponding recycling centres appropriately process those products. The chosen DPP implementation mechanisms should take this into account. In specific types of long-lived products (e.g., buildings, cars...) the data should remain available even when the manufacturer company has ceased to exist (e.g., because of bankruptcy). This is derived from Article 9.2.(h) in ESPR.

\item[R-05] \textit{Only specific actors, at specific stages, may \underline{create} a DPP or introduce  \underline{or update} any data elements in \underline{the DPP}}. This is a result of Article 9.2.(g) of ESPR. The design of the DPP system should specify which of the actors keeps the product-related data, as it has been already mentioned in R-02. Should it be a single actor or there may be some collaborative managers? Once this issue is solved, that manager or set of managers should also control which actors may create new DPPs or update the existing ones.

\item[R-06] \textit{\underline{Access to} different information items of the \underline{DPP} should be controlled and its permission will depend on the specific role of the requesting actor}. This is a consequence of Article 9.2.(f) in ESPR. Manufacturers, dealers, retailers, importers, customs officers, customers, end-users, repairers, recyclers may only access to subsets of the information present in a DPP. So, some (role-based) access control mechanism is needed to handle this. The rules to be followed will be stated in specific delegated acts for each product group.

%DIDs, combined with VCs and VPs, will provide an excellent basis to achieve that level of access control.

\item[R-07] \textit{DPP data \underline{authentication, reliability and integrity} must be ensured.} This is directly stated in Article 11.(g) of ESPR. So, DPP data should be only written by the intended agents, who should certify their identity in that DPP creation or update stages. Besides, DPP data, once written, should become immutable to ensure its integrity.

\item[R-08] \textit{DPPs shall be designed and operated so that a high level of security and privacy is ensured and \underline{fraud is avoided}}. Again, this is directly stated in the ESPR; in this case in point (h) of its Article 11. Fake DPPs must not be forged and accepted. Security and privacy were already partially addressed in requirements R-05, R-06 and R-07.

\item[R-09] \textit{DPP management should introduce, if any, a \underline{negligible burden} to all economic actors}. This is derived from Article 5.11 of ESPR. The technologies to be used should be open and should not introduce any noticeable costs that could endanger the competitiveness of the manufacturer or any other actors.

\end{enumerate}

All these requirements can be easily met when DIDs, VCs and VPs are jointly used to manage product identities and product certifications. That fact is analysed in detail in the subsequent section.

\section{Analysis} \label{sec:analysis}

% TO DO: At the moment this section provides only an outline of what should be explained here. We have to delve in each possible solution and describe carefully how they may be implemented using DIDs, VCs and VPs.

% TO DO: Besides the current section structure, we might need some preliminary sections that present the subsets of data in a DPP that may be of interest to each agent role. There is quite uncertainty in this regard, though, since DPP pilots have not clearly stated which is the information to be managed in different product groups (or product areas).

Each of the proposals presented in Section \ref{solutions} has made a set of design decisions, concerning decentralised identifiers and verifiable credentials, that lead to the implementation of digital product passports. This section focuses on those approaches, and complements them with other considerations to carefully analyse how the requirements presented in Section \ref{sec:reqs} could lead to different DID-based (or VC-based) designs for DPP management systems. 

% TO DO: We should review again our path of description in order to analyse whether the original decision trees may be recovered. It should, but that will not be immediate, unfortunately.
%This comparison aims to serve as a decision tree, as certain choices influence others. This tree will also help to locate current proposals and facilitate their comparison with potential future ones that may emerge.

\subsection{Product Identification (R-01)}
\label{sec:productId}

% One of the first and most complicated decisions to make in these systems is how products will be identified. 
As stated in Section \ref{normative}, ESPR currently mandates, in its Annex III, that products should use ISO-15459 identifiers, tolerating also other complementary identifications. Therefore, DIDs could be also added to a DPP, without replacing ISO-15459 identifiers, and such strategy does not endanger ESPR compliance. In spite of this, other research works have tried to integrate those ISO identifiers in DIDs, creating or suggesting specific DID methods to this end.

%fmunyoz: I think that the usage of digital twins (DT) does not impose any characteristic on the actual identification of products. DTs may be seen as a virtual entity that may support the physical product, but do not condition the identification strategy.

%The option that has emerged most frequently during the analysis of existing solutions is the use of digital twins. With these, a DID is generated for each unit or batch depending on the product's complexity.

As an example of the first approach, in \cite{Guth22}, it is shown how DIDs could be integrated with an identifier accepted in ISO-15459, such as %GS1. This solution is very specific because it is possible due to GS1 having a digital link that can be added to the DID document to relate both identifiers. However, not all types of identifiers in the ISO have this mechanism.
a GS1's digital link. Those links encompass a GTIN with some complementary information. Thus, that digital link may be added to the DID document to relate both identifiers: GTIN and DID. 

Similarly, another %way to create digital twins 
solution in this first approach
could be to include the ISO-15459 identifier as an additional attribute within the DID document, somewhat akin to the utility of the ``alsoKnownAs'' field defined in \cite{Sporn22}. Assuming a DLT network is used as a VDR, this solution would result in additional read load because finding the product's DID document from its ISO-15459 would require sequential reads until the correct one is found. Without a mechanism to facilitate this process, this solution would not be viable. However, other VDR implementations are acceptable, and they may introduce no penalty on read access time.

On the other hand, as an instance of the second alternative, Dr. Guth-Orlowski proposes %that another solution would be 
to encourage the various organizations that have standardised product identifiers (i.e., ISO/IEC and GS1) to create a DID method that includes the identifiers generated from each standard. For instance, 
%GS1 defines GTIN as the identity format for product labeling. Therefore, the organization responsible for managing the did:gs1:$<$gtin$>$ method, with DIDs in the form did:gs1:gtin, should be created. Conversely, if we follow the current regulatory proposal \cite{EC22}, 
the DIDs to be used in DPPs might become did:iso-15459:$<$iso-15459-id$>$. However, it is improbable that this solution will succeed because those organizations have displayed no interest in the implementation and use of DIDs. 
%Moreover, GTINs are already integrated into the ISO/IEC-15459 standard specification, so it would suffice to have a single DID method (the latter one mentioned) and a slight change in regulation \cite{EC22} (which could be limited to a brief paragraph, acknowledging that product identification could be done with both traditional standards and DIDs, specifying which DID method to use, or even leaving the method open, as interoperability between different DID implementations will improve in the future) to resolve the issue.

%The last option presented is the most different from the rest and represents a more radical change, which is to not identify products with DIDs. Initially, the W3C recommendation defines that DIDs can be used to identify any type of entity: individuals, companies, pets, or even objects. Despite this, DIDs are clearly designed for use by active agents who will have the necessary cryptography to authenticate and authorize relationships with other agents, as well as publish services in their DID documents. Conversely, products are passive agents that should be controlled by a person's or company's DID. This would make sense if there were no issues in using DIDs for products, but as seen before, this is not the case. Therefore, in the following sections, it will also be discussed what design changes would entail maintaining the ISO-15459 product identifiers and using only DIDs for active agents such as manufacturers, customers, and recyclers.

Finally, it is worth noting that, as an alternative approach, ISO-15459 identifiers could be the single way to identify products, since products are the unique passive entities in this CE ecosystem, while DIDs might be used for identifying the other entities (i.e., active agents) that intervene in the product lifecycle: manufacturers, dealers, customers, end-users, recyclers... Indeed, GS1 proposes in a recent white paper \cite{Dean23} that verifiable credentials should be used for certifying several product data with GTINs as the product identifiers in those certifications. This eliminates the need of associating a key pair for every identified product, since no DID document should be kept for them. However, that white paper also shows how to combine DIDs and GTINs, since VCs still use DIDs as the VC identifier while the GTIN identifies the intended product on which some certification is made.

Therefore, from the analysis of this first requirement, these design alternatives have been generated:

\begin{enumerate}
\item[D01.1]
\textit{DID and ISO-15459 identifier coexistence in DPPs}. VCs and VPs may refer to the DID embedded in the DPP. All three elements of SSI management may be used in DPP management.

\item[D01.2]
\textit{ISO-15459 identifier integration in DIDs, using the resulting DIDs in DPPs}. This choice preserves the functionality of the D01.1 alternative and, additionally, should not worry about identification duality, since the ISO identifiers are embedded in the DIDs. Its management should become easier than D01.1's one.

\item[D01.3]
\textit{Only ISO-15459 identifiers are used in DPPs}. Product identification is not based on DIDs in this alternative, but VCs and VPs may still be used, referring to ISO identifiers instead of DIDs. Since passive entities, like products, will not require many identifiers along their existence, but only a single one, no complication arises if VCs refer to ISO/EIC identifiers (e.g., GTINs). VC-based management is still possible in this alternative.
\end{enumerate}

\subsection{Support to Multiple DPP Granularities (R-02)}
\label{sec:granularity}

Article 9.2.(d) of ESPR identifies three different granularities for managing DPPs: model, batch or item. That means that for simple (or cheap) products that are usually discarded once they reach the end of their lifecycle, a single DPP per product model is enough: it will provide enough information (raw materials, weight, physical dimensions...) to deal with the product recycling and all items share the same characteristics. When some of those characteristics may change from a batch to the next, a DPP per batch will be needed. Finally, for more complex products, composed of multiple parts with different raw materials per part or able to be repaired in case of malfunction, a DPP per item will be used.

In general, in all DPP types, a DID (or even none if D01.3 is chosen) per DPP is needed. Besides that, these other aspects should be considered, leading to the following design choices:

\begin{enumerate}
    
\item [D02.1]
When the D01.3 design alternative has been chosen (i.e., no DID is used for product identification), products with a DPP per model or per batch will store the product information in a registry (e.g., a database) whose access might be controlled using VCs.

\item [D02.2]
In products identified with DIDs and with a DPP per model or per batch, product information should be stored in the DID document. Although the DID document is a structure that holds certain metadata of the entity (designed to maintain at least the public keys and access points to the corresponding service), the W3C specification \cite{Sporn22} allows for other information to be logically stored within it. To this end, those data that are not directly related to the standard document attributes, may be stored separately (as mentioned in \cite{Guth21}) and made accessible through the service endpoints kept in the document. Therefore, it would be advisable to establish what attributes could be included in that document regarding the product composition and the parts that can or should be recycled and the reasons for it.
    
\item [D02.3]
In products with a DPP per item, the composition of each recyclable piece should be documented using VCs, as suggested by Guth-Orlowski in her articles \cite{Guth21, Guth22, Guth23}. This requires the creation of specific VC schemas that describe the structure that VCs should have.
    % TO DO: This last item should be refined. There are some papers that provide examples of the data elements to be kept in a DPP, depending on the product group. Some of them, perhaps, could not be kept in VCs.
\end{enumerate}

\subsection{DPP Maintenance (R-03)}
\label{sec:maintenance}

% THE CURRENT CONTENTS SHOULD BE CAREFULLY REVISED. HERE, WE MUST DISCUSS WHETHER THE DPP MAINTAINER ROLE SHOULD BE EXCLUSIVE FOR A SINGLE AGENT OR MAY BE DELEGATED TO AN EXTERNAL SERVICE. FOR BOTH ALTERNATIVES, WE SHOULD CAREFULLY DESIGN HOW THAT RESPONSIBILITY MAY BE MIGRATED TO A DIFFERENT AGENT. WITH AN EXTERNAL SERVICE, ONLY THE REQUESTOR OF THAT SERVICE IS CHANGING IN EACH "MIGRATION".

%Apart from identifying the current owner of the product, it would be necessary to specify who will have the obligation to maintain the digital passport of a product. 
It is necessary to specify who is responsible for maintaining the digital passport of a product.
At the beginning of a product's lifecycle, the manufacturer will generate its DPP and logically, will be responsible for maintaining it. The diversity of options arises in the subsequent stages of the product's life. Once the product is sold for the first time, updating the owner within the DPP, as discussed in Section \ref{sec:dppUpd}, becomes necessary. With the change in product ownership, does it make sense to transfer control of the DPP to the current user of the product? This decision again depends on the nature of the product. If the DPP identifies a batch of products, such as plastic chairs, which are not expected to undergo repairs but are intended to go directly for recycling once damaged, having the manufacturer maintain the DPP would not involve any additional effort. If the information is registered in the DID document, it would be publicly accessible to all users and agents involved in the product's life cycle. In order to register changes within the DPP, the manufacturer could grant limited rights cryptographic keys, through which the agents involved in the product's life cycle can add specific information about their activity. 

However, if we consider a more complex product instead of a batch of simple ones, the scenario changes. In that case, a credential or a group of credentials may be used to represent the DPP, and that group would need to be accessible through a digital wallet. An example of a scenario of this kind arises when a person buys a bike, since it is normal to expect that during the bike lifespan, it will receive maintenance and possibly repairs. So, when the bike experiences any damage and requires, for example, the replacement of a part, the initial passport generated by the manufacturer will no longer correspond to reality. In that case, the bike remains the same, but its composition has changed, and this needs to be reflected in the passport. Once the manufacturer is no longer the owner of a product, they should not have to continue managing its DPP. It is not feasible for a manufacturer to maintain the DPP for all the products they produce. If a customer decides to change a product's parts outside the warranty period, that responsibility falls on them.

%Regardless of how the DID is represented, 
So, the most appropriate way to represent these product updates during its lifecycle will be through verifiable credentials. These credentials can be issued by an authorised technician or even by the user themselves. For instance, when a user takes their product to the relevant workshop for a repair, the workshop will issue a credential detailing the changes made to the product. At this point, the product's DPP will consist of a combination of the initial DPP generated by the manufacturer and the credentials the user holds in their wallet for that product. To achieve this resolution, two calls will be necessary. However, for a higher degree of decentralization, it's possible to provide the user with a copy of the DPP at the time of purchase to store in their digital wallet. This enables the entire resolution process of the DPP to be completed in a single call.

Thus, the design choices that arise in the analysis of this requirement are:

\begin{enumerate}
\item [D03.1]
\textit{No action}. When D02.1 or D02.2 choices have been adopted and the product is so simple that it should be recycled once damaged, no specific action is required to support R-03.

\item [D03.2]
\textit{Change of DID controller in DPPs supported by DID documents}. When D02.1 or D02.2 choices have been adopted and the product may change ownership along its lifecycle, a specific mechanism should be designed and implemented to change the DID controller entity.

\item [D03.3]
\textit{VC transfer in DPPs supported by VCs}. When D02.3 has been used, a specific mechanism for transferring VCs to a different product owner or manager may be needed.
\end{enumerate}

\subsection{DPP Data Availability (R-04)}
\label{sec:dppAvail}

In order to handle this requirement, two complementary facets should be discussed: (a) where to place the DPP data, and (b) how to handle its persistence along time. However, note that this requirement is implicitly met, since its solutions are directly derived from the design notes stated in R-01 to R-03.

\subsubsection{DPP Data Location}

DPP involves the gathering of information related to the product. The size of this collection varies depending on the complexity of the product. For simple products that do not undergo a complex recycling process and do not require repairs during their lifecycle, the passport will contain basic information such as the product's identity (serial number, model, description), manufacturer information, raw materials used in its composition, carbon footprint in its manufacture, carbon footprint in its usage, and relevant dates (production, sale, recycling). In more complex products, in addition to the aforementioned details, it is expected to include additional information such as warranties, repairs, certificates issued by external auditors, as well as the ownership history of the product. A complex product, in turn, may be composed of a combination of other complex and simple parts, and information about these should be referenced in the final product's DPP.

As explained, the DPP for simple products will have a shorter length. 
%Assuming that products have digital twins, 
In that case,
the information about their DPPs can be recorded in the DID document itself. Alternatively, it can be presented in the form of a verifiable credential issued by the manufacturer at the time of production, with the ISO-15459 identifier as the subject of the credential. Normally, verifiable credentials contain two DIDs: the issuer and the subject. The W3C recommendation allows for the possibility of omitting the subject DID in scenarios where the credential controller is irrelevant, and it is only necessary to know who issued it and what information it contains for verification; this is known as a bearer credential.

In the case of complex products, the situation changes. The option of using the DID document to record DPP information implies that this document should reference the rest of the documents for the components that make it up, potentially forming a tree with several levels of depth. This would result in as many blockchain calls as nodes in the tree, considering the slow reads in this type of network. For example, a bicycle can be composed of around 250 pieces depending on its complexity. For this reason, this option may become impractical. On the other hand, the option that has appeared most frequently in the analyzed solutions is the idea of forming the DPP as a collection of verifiable credentials. These would be stored in the owner's wallet, making the passport inquiry a single call to their wallet. The credentials forming the DPP can be of different types, presenting different information schemes for each scenario. Within the collection, you may find the initial product credential, credentials about possible repairs, maintenance, or previous purchases, as well as a list with the DPP of the components that make up the product.

\subsubsection{DPP Data Persistence}
\label{sec:persist}

Simple products may keep their DPP information in their DID documents, since in the regular case that information does not need to be updated or extended along time and may be shared by all the manufactured items with the same model or product reference. Note that DID documents are kept in the VDR and that VDR is immutable in most DID systems. Therefore, there will be no trouble to ensure data persistence in that case.

% TO DO: This should be revised carefully. VCs transfer should be explained in detail in a separate section. That can be one of our contributions!
On the other hand, complex products may raise some difficulties, since they cannot hold their DPP data in a DID document. At a glance, the most realistic choice is to keep those data in multiple VCs, one per kind of information. Those VCs have been generated by the manufacturer and they will eventually be kept by the product owner, in its digital wallet. If the product is sold, those VCs need to be transferred to the new owner. Similarly, at the end of the product lifecycle, a VCs transfer or final conversion into a set of VPs to the recycling centre will be applied. Therefore, VCs transfer should be implemented in order to manage this kind of DPPs.

\subsection{DPP Creation and Updating (R-05)} \label{sec:dppUpd}

% fmunyoz: This section has to be revised in regard to access control, at least, in simple products that have a single DPP per product model. In that case, regular customers may read some parts of the DPP data, while other roles, e.g., the recycling centres, may be allowed to read other parts that are kept secret to customers. Who controls data access? In those simple products, at a glance, it seems recommendable to assign that controller role to the manufacturer or to any external agency agreed in that product area.
%Once the product is identified, it is necessary to determine its owner, which will be crucial in the following sections to define who will be responsible for maintaining the DPP or how information can be added to the DPP. 

Identifying products with DIDs offers certain design advantages in regard to product-associated data insertion or updating, as the W3C DID recommendation \cite{Sporn22} has several sections dedicated to this specific scenario. A DID can refer to a subject but be controlled by another. In this case, a product would never control itself; there must be an owner at all times who carries out interactions on its behalf. For this reason, the DID document supports a 'controller' attribute, which defines the controller of the said DID. The cryptography made public in the product document would be generated and controlled by its owner. Initially, this would be the manufacturer. 
% What is mentioned in the next sentence may depend on the type of product. So, we must carefully extend our explanation in this regard.
Subsequently, when a change in ownership is needed due to the product sale, simply updating the document by overwriting the 'controller' field with the new owner's DID would suffice.

%As mentioned in the previous section, 
However, identifying products without DIDs brings about certain issues, so let us discuss what it would mean to not use them for product identification. Without a DID, the mechanisms defined by the W3C mentioned above are lost, so new mechanisms must be established through verifiable credentials. This alternative would involve the use of possession credentials. At the beginning of a product's lifecycle, the owner would implicitly be its manufacturer. At the time of sale to a new owner, the manufacturer would issue a possession credential, which would have the new owner's DID as the subject of the credential and the ISO-15459 identifier of the product just acquired as an attribute of the subject. This information will be attested to by the signature the manufacturer adds to the credential.

This design is akin to using purchase receipts in the physical world. It would simplify interactions, such as requesting a product warranty. In this scenario, the owner would create a verifiable presentation of the credential issued by the manufacturer at the time of sale. However, in a more realistic scenario, a product will pass through many owners during its lifecycle. After its initial sale, a product might be gifted or designated for resale. In these situations and others of a similar nature, the transfer of product ownership necessitates the generation of another possession credential by the current owner for the subsequent owner. This new credential would include the previous credential issued by the manufacturer, effectively creating a chain of easily verifiable credentials.

As we have seen, R-05 may be met reusing the tools available for supporting previous requirements. Nothing else needs to be considered here.

\subsection{DPP Access Control (R-06)} \label{sec:dppAccCtrl}

In the general case, a DPP stores different kinds of information and each kind may be only of interest to a few agent roles. Perhaps, in some cases, some of those data should not be read by some specific agent roles. Thus, the DPP maintainer, if any, should take care that each agent role can only access those data it is authorised to. To this end, VCs may be provided to interested readers, stating the role they may play in the DPP system and providing permission for some specific types of access on the interesting data.

For instance, a customer may be interested in some of the raw materials a product is composed of, e.g., in specific allergens. DPPs may include a complete list of the product's raw materials, since that information may be required at the recycling stage, not being available to regular consumers, though. However, if any of those raw materials is an allergen, then its existence in the list of materials will be readable by every role, even for customers that have not yet purchased the product.

So, let us analyse how to manage different categories of information depending on its availability (e.g., publicly available, available for specific roles, or only available to the owner) and which access controlling mechanisms may be used to handle them.

Information publicly available may be kept in locations (e.g., product documentation included in the product package, or in a website) that do not require any specific access control mechanism. % TODO: To be extended.

Information available to specific roles may be placed in a location that requires electronic access. Such access may be controlled requiring that the specific role is proven using a VP that shows the ownership of a given VC stating that such an agent belongs to that requested role. Thus, the service that controls those accesses behaves as a credential verifier and will accept those attempts when they carry the required VP. % TODO: To be extended, describing a specific example.

Therefore, the mechanisms that control product information accesses may be VCs and VPs, since they are able to handle such management. However, nowadays, it is impossible to place any design choices or considerations in R-06, since ESPR Article 9.2.(f) states that the security policies to be followed should be stated in future delegation acts and will depend on the concrete product group being considered. Because of this, no concrete design choice may be defined yet.

\subsection{Data Authentication, Reliability and Integrity (R-07)} \label{sec:dataIntegrity}

DID-based systems use a VDR \cite{Sporn19,Sporn22b,Sporn23} to keep some information, like DID documents, for instance. Blockchains may be used for implementing VDRs. Thus, VDR immutability and integrity may be easily ensured. If DPPs are implemented using DIDs, some invariable identity-related attributes of a DPP will be kept in the VDR (or in an external database that synchronises the DLT contents~\cite{Gonza23}, if read access time would become an issue), since they could be stored in DID documents. 

Those other parts of a DPP that could vary along time may be written and handled as VCs. VCs are not kept persistently in the VDR, but they are delivered to the interested agents, who keep them in their digital wallets. But, again, that information becomes immutable in its validity interval (note that VCs have an expiration time or date) since the corresponding data structure has been digitally signed and any tampering attempt will be easily detected. Besides data integrity, those signatures also ensure that the information has been generated by the intended party and that the protocol that handles such information is reliable.

A third type of information that may be of interest in the DPPs of some specific product groups is that related to traceability of relevant supply chain events. This kind of data will be implicitly stored in DPPs if the DPP of a complex product hierarchically (i.e., recursively) includes the DPPs of all its components and each DPP stores an ordered list of the events that have generated that item.

Thus, all aspects considered in this requirement can be successfully met with DIDs, VCs and VPs. So, no design choices arise here.

\subsection{Fraud Avoidance (R-08)}
\label{sec:fraudAvoidance}

Most industrial products use GTINs as their identifiers and those GTINs can be easily seen using the appropriate data carrier (in many cases, a barcode in their label). With GTINs, product identification becomes trivial, but, unfortunately, its support for preventing frauds from happening is too limited: that barcode can be copied and attached to a fake clone. Depending on the quality of that clone, it might be impossible to detect that fraud.

DPPs should be able to avoid those fraud scenarios. To this end, DIDs, VCs and VPs may be helpful. Thus, each time a new stage is reached in the product lifecycle (mainly, in the initial ones from its manufacturing, distribution, storage at the retailer, and sale to its first end user), that transition could be recorded generating a new VC that states the transition date and time, with the involved parties. That VC will be presented as the input for the next stage, where another VC will be added, generating a chain of VCs that records all those events. If all those agents identities and public keys are maintained at a given commercial VDR (i.e., a public registry that contains the name and public keys of all legal commercial companies, e.g., suppliers, manufacturers, distributors, retailers...), the authenticity of those steps might be verified by the customer at purchasing time. In that way, no fake product had any opportunity of confusing the customer, since those involved fraudulent actors would not be able to present a valid signature in their claimed VCs.

In that scenario, a key element is to compel every agent to create and sign a new VC that includes all the existing path of previous stages/VCs. Thus, if a fraudulent retailer FR tries to impersonate a legal one LR, FR will not be able to do so, since it does not know LR's private key for signing that final VC. Moreover, if FR signs that VC with its own FR (private) key, that fact will be also detected by customers:
\begin{itemize}
\item
FR's identity and key will not be in the commercial VDR previously mentioned. Thus, the VP to be presented to the customer will not have any verifiable signature since FR does not know LR's private key. Therefore, FR uses its own private key for signing the VP and its corresponding public key will not be in that public registry. 
\item 
In the previous product lifecycle stage, the generated VC had a given subject that would not match FR's identity, but LR's one.
\end{itemize}

A system that follows a similar approach (in the sense of using a chain of related VCs) has been designed and presented (although it is still a work in progress) by the W3C \cite{Otto24}. In its Appendix B.2, that working draft explains a use case where VCs provide claims about companies and products identified with GS1 keys (e.g., product identification is managed with GTINs). Since every VC may be identified using a DID, the next VC refers to the previous one using its DID. In that way, a chain of related VCs may be easily implemented.

Thus, to appropriately handle R-08 a new design note should be provided:
\begin{enumerate}
\item [D08.1]
\textit{A chain of related VCs is needed}. Each element in that chain certifies who and when generated a transition in the product life cycle.
\end{enumerate}

\subsection{Negligible Burden to Actors (R-09)}
\label{sec:noBurden}

Article 5.5 of the ESPR mentions that all DPP management should have no negative impact on the competitiveness of the involved actors (mainly, manufacturers, dealers and retailers), nor impose any proprietary technology, nor introduce any administrative burden for those actors. 

Although DID-based identity management is a new technology that has not yet had a wide adoption, it may help in this regard. To begin with, the recommendations published by W3C about DIDs, VCs and VPs \cite{Sporn19,Sporn23} admit and promote open implementations. Indeed, there are several libraries and tools available that facilitate those implementations with no licence fees. Once accepted and adopted by the general public, there will be personal digital wallets that will make VC- and VP-based interactions very easy. Besides, the involved product-lifecycle-related companies already handle all the information that should be present in DPPs. So, storing also that information in the required resources for implementing that DID-based system will not become expensive nor time-consuming.

No additional design consideration is needed to handle R-09.

\section{Proposals} \label{sec:proposal}

%TO DO: THE ANALYSIS SECTION MUST IDENTIFY THE MECHANISMS OR PROCEDURES TO BE IMPLEMENTED IN ORDER TO SUPPORT AN EFFICIENT AND RELIABLE DID-BASED DPP SYSTEM. THIS SECTION SHOULD REFINE THAT ANALYSIS, PROVIDING A DETAILED DESIGN THAT WE WILL IMPLEMENT.

%SECTION "6. SCENARIOS" SHOULD DESCRIBE SEVERAL USE CASES AND EVALUATE THAT IMPLEMENTATION IN THEM.

Let us briefly discuss how to implement each one of the design alternatives identified in Section \ref{sec:analysis}. If any of them requires a thorough discussion, that explanation is given in a subsequent part.
 
%DID and ISO-15459 integration (D01.2) has already been discussed and implemented in other papers like \cite{Guth22} (with its \texttt{iso-15459} DID method) and \cite{Navar22} (with the \texttt{eCHID} DID method), outlined in Sections \ref{solutions} and \ref{sec:productId}. Therefore, that line has already been explored and is not a central piece of our proposal.

The three design alternatives (D01.1, D01.2 and D01.3) related to product identification (R-01) have already been described in detail in Section \ref{sec:productId} and do not require specific mechanisms to implement them. Implementations of D01.1 and D01.2 have already been given in related papers \cite{Guth22,Navar22}, while the convenience of D01.3 has also been thoroughly explained in \cite{UNECE22}. Similarly, the design choices outlined in Section \ref{sec:granularity} in regard to DPP granularity (R-02) have implementations in other papers such as \cite{Hoops24} (since it describes how to use VCs to control the access to a system, as required in D02.1), \cite{Navar22} (for D02.2) and \cite{Guth22,Guth23} (for D02.3). Similarly, option D03.2 (change of a DID controller) is an issue that has already been discussed in Section B.12 of the DID recommendations \cite{Sporn22}, while D03.3 (VC transfers in DPP) becomes a subcase of D03.2 when a DPP supports jointly DID and ISO-15459 identifiers, as suggested in D01.1, since in that case each product will have its own DID and the current product owner will be the DID controller. In case of assuming D01.3 (i.e., no DID per product), then D03.3 could be implemented using a non-fungible token (NFT) per product, modelling product ownership with the NFT ownership. DPP data availability (R-04), DPP creation and updating (R-05), DPP access control (R-06), data authentication, reliability and integrity (R-07) and negligible burden to actors (R-09) are intrinsically handled in a correct way when VCs and VPs are used. Therefore, they do not need any specific proposal for their management.

So, only the mechanisms discussed in Section  \ref{sec:fraudAvoidance} (R-08) demand specific proposals to be described subsequently. Those descriptions will also analyse which combinations of solutions to the other requirements make sense.

\subsection{Management of Verifiable Credential Collections}

A DPP will contain relevant information about the product, such as the raw materials it is made from, data obtained during manufacturing such as CO$_2$ produced, or certificates issued by other organizations. In a system based on decentralised identity, all this information must be contained in the form of verifiable credentials, so the DPP will not be a simple document but a collection of these credentials. We will refer to this collection generated by the manufacturer as the original DPP, as it represents the state of a product that has just begun its life cycle and has not yet received updates.

In the case of simple products or batches of complex products with the same manufacturing process, the DPP will be identified by the GTIN as a class or product identifier. The original DPP with only the GTIN will be sufficient for simple products that will only be used to form part of other more complex products or for objects that, once broken, will go directly to recycling without receiving repairs. For complex products like bicycles or cars, it is expected that they will receive updates such as part replacements or inspections during their life cycle. This makes it necessary for the DPP to be extensible and for mechanisms to be defined that allow different agents to add information about a product during the different stages of its life cycle. From the moment of manufacture, the credentials generated to extend a product's original DPP will refer to an instance of that product, so in addition to the GTIN, its serial number must be included.

Using vehicles as an example of complex products, it is expected that the agents responsible for issuing information about the vehicle after its manufacture will mostly be the workshops that the owner hires for maintenance. When a workshop performs work, it must issue a new credential to the owner, containing information about the task performed or the changes made concerning the original DPP. This credential must contain both the product's serial number and the GTIN to be related to its DPP, becoming part of the collection. Although we refer to all credentials as a collection, they are not stored together. The original DPP is stored by the manufacturer, and the owner keeps updates about an instance of the DPP.

The issuance of this credential will have been done privately between the owner and the workshop, but this extension of the DPP must be published in some verifiable data registry to achieve product traceability. This is necessary because the newly generated credential will be stored in the owner's digital wallet, but the owner might decide not to share this information with other workshops or potential buyers.

The process of resolving a DPP based on credentials and DIDs is well defined in Dr. Guth-Orlowski's proposal. In this proposal, from the DID of a product, one could query its document, which must have an address to the manufacturer's digital wallet to request the DPP. This solution is sufficient to obtain the DPP of simple or complex products that have not received updates.

As explained earlier, the original DPP will be generated by the manufacturer and stored and made available for possible queries by other agents. Although, at the time of purchase, the new owner could store a copy of that original DPP in their digital wallet, it is not necessarily mandatory as it would duplicate the information.

To perform the process of querying the current state of a certain product's DPP, one would need to obtain the original DPP identified by the GTIN, either from the manufacturer or the current owner, and apply in order the updates that the owner might have in their digital wallet in the form of verifiable credentials. All credentials must refer to the same serial number.

\subsubsection{P08.1: VCs and DID per Product} 

\begin{figure}[h]
\centerline{\includegraphics[width=0.8\textwidth]{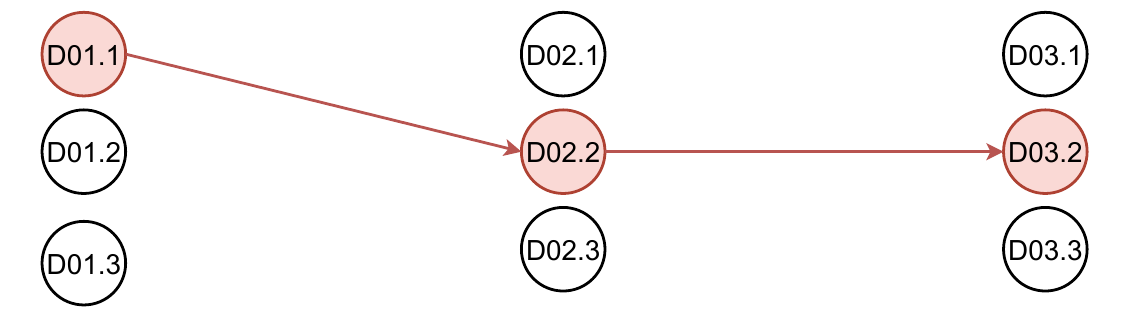}}
\caption{Design options in P08.1.}
\label{fig:grafo1}
\end{figure}

Working only with credentials results in a lack of traceability, as they are issued and stored privately between the two interacting agents. To address this, mechanisms are needed to identify both the agents and the product and to register the events that occur during their interaction. A product, in addition to having a collection of credentials as its DPP, should also have an associated DID. This new identifier is not intended to replace the current product identification system (GTIN + SN) but to complement it by enabling a digital twin for each product. The product will be identified in the corresponding DID document, thus allowing the relationship between the DID, GTIN, and serial number. By extending the functionalities of DID documents, it could be defined that the controller of the document corresponds to the current owner of the product. Design options chosen for this proposal are represented in Figure \ref{fig:grafo1}.

Unlike credentials, DID documents will indeed be registered in some public registry. To check who the owner of a product is at a certain time, it will only be necessary to read the \textit{controller} attribute of the document. Identifying the owner publicly would pose a security and privacy issue if personal data such as a national ID or Social Security number were used. For this reason, the owner should be represented in the document by an anonymous DID that they should only use for their relationship with that product. The fewer DIDs are reused for different use cases, the harder it will be to discover the identities of the subjects of those identifiers through DID correlation.

\begin{figure}[h]
\centerline{\includegraphics[width=0.9\textwidth]{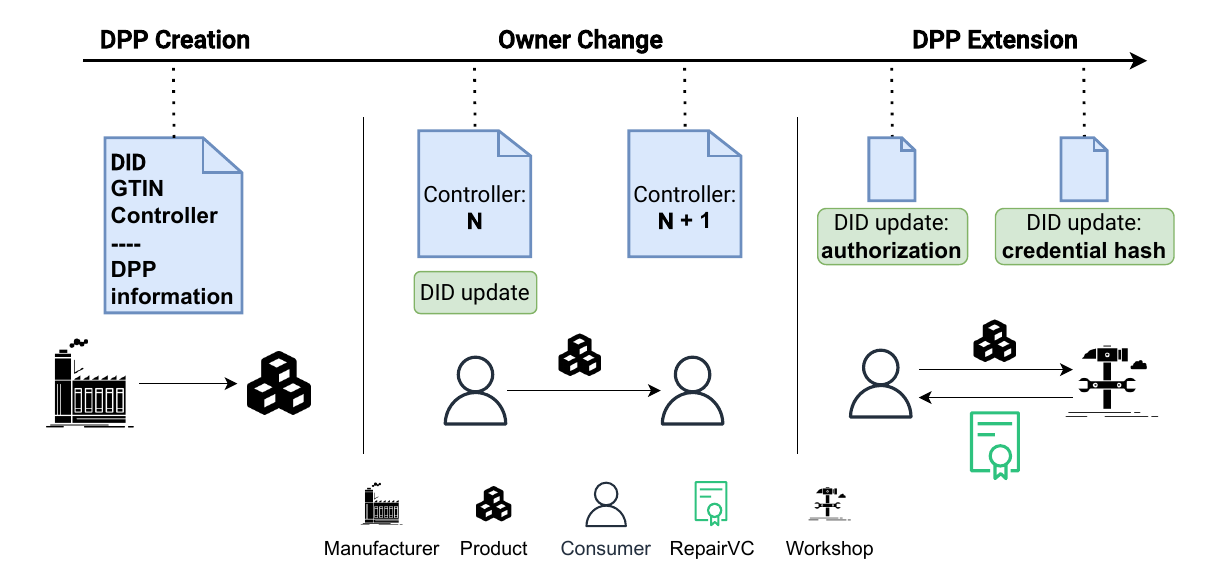}}
\caption{DPP lifecycle representing the events when the DID information is updated and when a VC is used, in three different scenarios: product creation, transfer, and information extension.}
\label{fig:dpp}
\end{figure}

When the product is sold to another agent and the ownership change needs to be registered, the current owner ($P_n$) only needs to replace the controller field of the document with the DID of the new owner ($P_{n+1}$), as shown in Figure \ref{fig:dpp}. With this change, $P_n$ will no longer have control over the document and will not be able to modify it. $P_{n+1}$ will need to update it to add their own public keys. In most cases, the DID document will be registered on a DLT-type network, so in addition to being able to consult the current controller, it will also be possible to access the entire history of owners that the product has had.

In the example of DPP extensibility, the need to publish the process in a registry to achieve the traceability of products and their passports throughout their lifecycle was mentioned. To implement these mechanisms, the functionalities of DID documents can be employed. Initially, the controller of the product will be the only one able to make updates to the document, but the DID specification defines the use of delegation keys. These keys would allow another agent to update the document.

Returning to the example of the vehicle workshop, once the repair is done, the workshop will issue a credential with the information to the owner. Additionally, the owner must grant permission to the workshop to update the document and register this event using a hash of the credential. This will allow maintaining the traceability of a product while keeping the information in the credentials private. Furthermore, registering the credential's hash in the document will serve to verify that the owner is not attempting to hide information related to the product in situations where this information is relevant.

\subsubsection{P08.2: VCs per Product}

The previous proposal was based on the use of verifiable credentials to register product updates and DIDs to identify both the agents performing these updates and the products receiving them. The alternative presented in this section aims to address the same issue but using only verifiable credentials in all possible aspects, thereby eliminating the need to identify the product with a DID or to use the functionalities of the DID document.

W3C establishes that the subject of a verifiable credential does not necessarily need to be a DID, as long as it can be universally identified through the rest of the data. Based on this, a product's credentials do not need to refer to its DID, but to its GTIN and serial number. This way, the need to generate an additional identifier for each product instance is eliminated. However, this does not imply the complete elimination of DIDs, as they are still necessary to identify the other agents in the chain.

\begin{figure}[h]
\centerline{\includegraphics[width=0.8\textwidth]{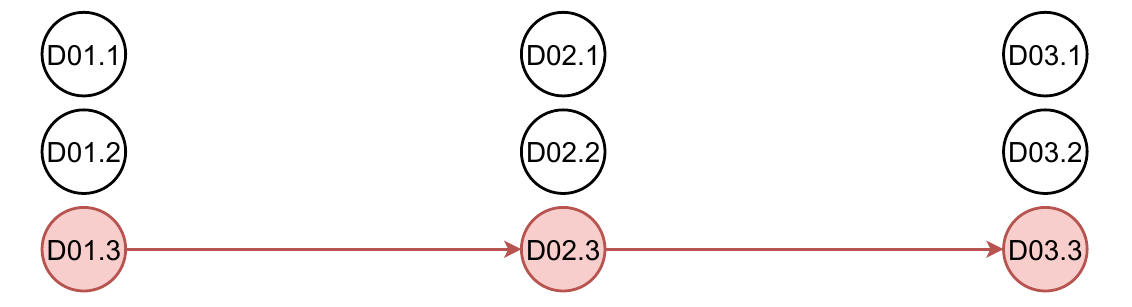}}
\caption{Design options in  P08.2.}
\label{fig:grafo2}
\end{figure}

By dispensing with a DID, access to the associated document that specifies the controller's DID is lost. Therefore, an alternative that allows the owner to prove possession of the product must be considered. A viable solution would be to implement a ``transfer credential'', a verifiable credential that functions similarly to a physical purchase receipt. This credential would be issued by the seller to the new owner and stored in their digital wallet.

In the event of a future resale, the current owner would issue a new transfer credential that includes the original receipt credential obtained from the seller and the current transaction information, such as the date of the sale, the price, and the new owner. This new transfer credential would be given to the new owner, who would store it in their digital wallet. In this way, a chain of transfer credentials would be created, tracing back from the current owner to the manufacturer, including all previous owners and the store where the product was originally sold. This chain of credentials would serve as proof of possession and allow tracking the product's ownership history.

\begin{figure}[h]
\centerline{\includegraphics[width=0.9\textwidth]{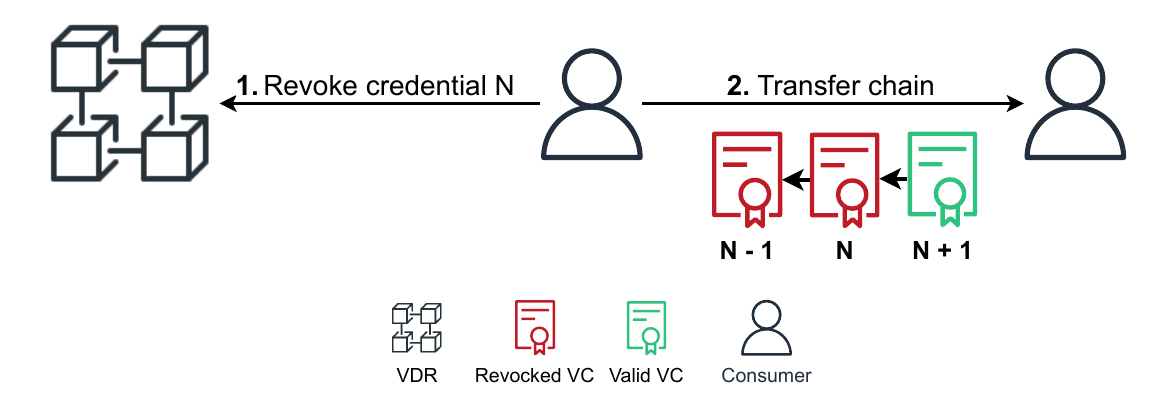}}
\caption{Steps for the transfer of control of a DPP through a chain of verifiable credentials, where the last one represents the current owner.}
\label{fig:proposal2}
\end{figure}

As in the previous proposal, it is crucial to ensure that the change of ownership is securely recorded. Since the issuance of the transfer credential is a private communication, the previous owner could still present their previous credential, which could be successfully verified. For this reason, during the issuance process, it must be ensured that the seller revokes their current transfer credential so that the only valid one is the buyer's. Figure \ref{fig:proposal2} illustrates the steps in this process.

An alternative to revocation by the owner would be to request that the manufacturer revoke the current credential and issue a new one for the new owner, acting as a trusted provider. While this option is simpler, it presents some disadvantages: manufacturers do not have an inherent interest in maintaining a long-term relationship with the owners, as their interest ends once the product is sold. To address this, incentives could be established for manufacturers to maintain the service, but they could not be forced. Additionally, the main problem lies in the dependence on the manufacturer as a trusted provider, which contradicts one of the primary goals of using DIDs and verifiable credentials: eliminating intermediaries.

Proposal P08.1 explains why it is necessary to register credentials in the DID document. In P08.1.2, since the product is not identified with a DID, there is no document available for this registration. Therefore, the implemented data registry must be used to register the hash of the credentials. In the case of a DLT, this would be done through a transaction that includes the product identifier and the hash of the issued credential.

\section{Discussion} \label{sec:discussion}

This section discusses the advantages and disadvantages of the two previous proposals. A preliminary analysis of the implications and the feasibility of each proposal presented in the previous section will be conducted.

\subsection{Pricing}
\label{sec:price}

One of the most important factors when evaluating the proposals will be the cost for the users of the system. This cost will primarily depend on the DLT network chosen for implementation. Public networks like Bitcoin or Ethereum have transaction costs ranging between two and seven euros, whereas other layer 2 solutions can significantly reduce this cost to just a few cents.

Beyond the choice of the network, the cost will also vary depending on the specific proposal implemented. This is because each approach has different operational needs within the DLT network. On the same network, the cost of these operations can be up to 20 times higher depending on the selected proposal.

For this comparison, we have chosen the Cheqd network as an example. Cheqd \cite{Cheqd24} is a DLT specifically designed to support decentralised identity and, at the same time, has significantly lower costs than some of its alternatives. A clear example is the deactivation of a DID: on the Sovrin main network, this process costs 10 euros, while on Cheqd it only costs 0.40 euros. On Cheqd, the prices of operations are tied to the price of the network's cryptocurrency, so the cost will be approximate and could vary.

The choice of the DLT network and the specific proposal will have a significant impact on the operational costs for the system's users. Therefore, it is crucial to conduct a detailed analysis of these aspects during the proposal evaluation phase. Finally, the cost analysis of the proposals will be calculated from the perspective of the product owners, as they will bear most of the system's costs.

In proposal P08.1.1, each product will need to have its own DID. This DID will be defined at the beginning of the product's lifecycle. Therefore, the manufacturer will have to assume its cost as an additional manufacturing cost. On Cheqd, the cost to register a DID is approximately 50 tokens of the network, and during the DID registration process, Cheqd allows the use of a custom DID document, so there would be no need to update the document after the DID is registered. On the other hand, to be able to control the identifiers being generated, the manufacturer will need their own DID. Subsequently, once the product leaves the factory and is sold to another agent, the manufacturer will need to update the document to transfer its control to the new owner, which would cost 25 tokens. Therefore, the cost for the manufacturer would be the following equation where \textit{x} is the number of products and \textit{t} is the value of the tokens:
\begin{equation}
    50t + 75xt
\end{equation}
The equation represents the cost of the system for the manufacturer, as they will first need a DID, and from that point on, it will be 50 tokens for the product's DID, plus another 25 tokens when the product is transferred.

At the time of writing this article (i.e., June 2024), Cheqd tokens are priced at 0.04 euros, so apart from the company’s own DID, which would cost 2 euros, each product would have an additional cost of 3 euros. After the manufacturer transfers the product, subsequent owners would only have to pay 25 tokens, equivalent to 1 euro, for each DID document update. This would occur whenever the product undergoes an event, such as repairs or inspections, or in the case of a change of ownership.

With proposal P08.1.2, products would not need a DID, and therefore owners would not have to pay the cost of updating the DID document. In this system, based primarily on credentials, the cost is significantly reduced, as both the price of creating a status list for the credentials and updating the list to indicate their revocation is only 2.5 tokens (0.1 euros). The following equation represents the cost for manufacturers, including the cost of an initial DID, the issuance of the passport, and the issuance of the possession credential.
\begin{equation}
    50t + 5xt
\end{equation}
We say ``based primarily on credentials'' because both the manufacturer and the other agents participating in the product's lifecycle will still need at least one DID. Although the cost of DIDs is present in both proposals and one could argue that it is not necessary to mention it, it must still be included, as it is an additional cost associated with using DIDs and verifiable credentials.

\subsection{Traceability}
\label{sec:trace}

Proposal P08.1.1, based on DIDs, offers superior traceability by identifying products with a DID and recording their events in the associated DID document. This allows any agent with the product's identifier to verify its ownership history and received events. Publishing the ownership history and events in the DID document might seem like a loss of data privacy, but it is important to consider that the owners are not identified with personal information but with an anonymous identifier, which, for greater privacy, will only be used in relation to the product, thus preventing correlation.

Detailed information about the events is stored as verifiable credentials in the current owner's digital wallet, protecting privacy and preventing unauthorised access. Unlike verifiable credentials, which require each issuance to be recorded on the blockchain, the DID document allows for a more seamless and straightforward query of the product's history. This is due to the centralization of information in one place. To achieve similar traceability in proposal P08.1.2, each issuance would need to be recorded on the blockchain, which is more complex. Moreover, querying the complete history would require retrieving and verifying each transaction individually, making the process slower and less practical.

Proposal P08.1.1 offers significant advantages in terms of product traceability, providing a transparent history, data protection, and efficient queries. While verifiable credentials offer greater privacy, their use for traceability is less efficient and scalable.

\subsection{Chosen Proposal}
\label{sec:chosen}

After analyzing both proposals in detail, it is time to evaluate which one is more suitable for implementation.

One of the main requirements of the DPP is that the implemented system should not impose excessive costs on manufacturers, as this could jeopardise their competitiveness in the market and lead them to avoid its implementation.

% fmunyoz: Let us rewrite the first sentence of the next paragraph, with a more direct style.
%In section 6.1, the cost of each proposal was discussed, and it was found that the difference is considerable, with P08.1.1 potentially costing around 15 times more than its alternative. 
Section \ref{sec:price} has discussed the cost of each proposal, and it has found that the difference is considerable, with P08.1.1 potentially costing around 15 times more than its alternative. 
Despite this significant difference, if the network for implementing the system is carefully chosen, this difference will amount to only a few cents or a couple of euros.

Considering these costs for complex products like a bicycle, whose average price is around 1,200 euros, an additional couple of euros for enabling the DPP can be considered negligible. However, for simpler products with much lower prices, such as a light bulb, a couple of euros could represent a 50\% increase in its price. This would be the case if light bulbs required a DID for each product instance, but this is not the case, as simple products are recognised because they will not receive updates such as repairs; if a light bulb stops working, it is not taken to a workshop, it is simply recycled. Therefore, for simple products, the cost would be even lower, as only one DID per product type would be needed.

Subsequently, in Section \ref{sec:trace}, the advantages of traceability and information availability offered by proposal P08.1.1 compared to P08.1.2 were presented. After analysis, we consider that these advantages outweigh the cost difference between the proposals.

\section{Use Cases} \label{sec:use_cases}

To better visualize the use of the digital product passport along with verifiable credentials and decentralized identifiers, we present two different scenarios. The first scenario considers a simple product that is not intended to receive repairs, as it is better to recycle it and purchase a new one once it breaks. The second scenario presents a complex product that contains certain electronic components which, in case of failure, could be replaced to restore the product's functionality.

\subsection{Simple product}

For the first product, we have chosen cycling glasses. These consist of two components: a lens, which is a single piece, and the frame. Unlike the lens, the frame is not a single piece as it is composed of four other parts. The frame of the glasses is formed by joining the temples, the temple tips, the lens frame, and the nose bridge pads. If the lens of the glasses breaks, it would not make sense to buy another lens to replace it, as the price would be very close to buying new glasses. For this reason, this type of simple product that is not expected to receive updates does not need a digital passport per product instance, but rather per reference.

As mentioned, cycling glasses consist of two distinct parts: the frame and the lenses. It is easy to assume that both parts will be manufactured in different factories and then assembled together at another location. At the time of their manufacture and departure from the factory, each part must have its own digital product passport (DPP).

For the frame, since it consists of a simple product and in case any part of the frame breaks, the easiest solution would be to buy a new frame or new glasses altogether, the DPP could follow the design option D02.2. With this, all the information about the frame’s parts will be available in the document of the DID that identifies that model of frame. In the case of the lenses, the DPP will be a DID document containing their information, such as the type of lens used (i.e., its material composition), its UV filter grade, and the ISO standards it meets.

\subsection{Simple electronic product}

In the second case, we have selected a computer mouse. This device, although it may seem simple at first, contains a series of fairly complex electronic components that lend themselves to repair in case of failure. This scenario will not apply to all mice, as a wired mouse purchased for 10 euros, no matter how simple the repair, would exceed the purchase price. This is not the case with more modern and complex devices, such as some models that exceed 100 euros and, besides containing more electronics and components, also include batteries in the case of wireless models.

Despite containing many electronic components inside, a mouse can be divided into two groups of parts: the casing and the motherboard. The casing, in most cases, will be made of some type of plastic or rubber and will consist of the various buttons that the mouse contains and the larger parts that connect them. However, similar to the frame of glasses, it will be a very simple product passport whose information will mainly serve to indicate what type of plastic has been used and how to recycle it. For this reason, its DPP will be the same as that of the frame of the glasses.

In contrast, the motherboard will be a collection of the passports of many electronic components such as sensors, LED lights, microchips, and even batteries, which already contain a greater variety of materials in their composition such as metals and silicones, requiring more specific handling for their subsequent recycling. In total, the motherboard may have more than 10 different components that may be subject to repair in case of failure, so each of these may have its own passport. Here, the design option D02.3 may be the most appropriate, since in case the battery fails, its credential within the passport will be replaced by the credential of the new battery.

\subsection{Complex product}

The use cases that have been analyzed correspond to a fairly simple product that is not expected to receive any repairs, and another product of reduced complexity but already featuring electrical components that will require more customized recycling and that, in certain cases, could indeed be subject to repairs. Finally, to better visualize the needs and limitations of the digital passport, we will present another use case, one that represents more complex products due to their number of components and the amounts of repairs and maintenance they will receive throughout their lifecycle. A car is a product that fits perfectly with the needs we have described.

In detail, a car can have between 70,000 and 90,000 different parts. If we only used credentials to store the information of all the components, the collection of credentials that will form the product's digital passport and that the owner will need to keep in their wallet could take up from 300 megabytes to nearly half a gigabyte of storage. This represents a considerable burden for the user, and the implementation of the passport should be as transparent as possible for both users and companies. Due to this, another implementation needs to be considered for these highly complex products. To attempt to reduce the size, the use of credentials could be minimized. In the case of cars, instead of being a collection of credentials, the passport could be a DID document, which references its most direct components, whose passports would point to their own components, and so on, down to the simplest parts. This type of passport could be generated by car models or batches. Then, the information on repairs for a specific car instance would be recorded in verifiable credentials in the owner’s wallet. With this model, the owner would handle a significantly reduced number of credentials compared to the first option we presented for the car's DPP.

\section{Conclusion} \label{sec:conclusion}

This article presents an in-depth analysis of the European Commission's Digital Product Passport (DPP) proposal, a critical initiative for advancing the circular economy in the region. The DPP emerges as a key element in driving transparency, traceability, and sustainability across supply chains, promoting more efficient resource utilisation and waste reduction. The primary objective of this article is to demonstrate the viability of using decentralised identity (DID) for implementing the Digital Product Passport. To achieve this, existing SSI-based solutions are thoroughly examined, exploring their potential application within the DPP context. While these existing solutions present valid proposals and showcase the ability of DID and verifiable credentials to support specific passport scenarios, limitations in their scope to encompass all stages of the product lifecycle are identified. Instead, the proposal presented in this article defines mechanisms to accompany the product throughout its useful life, supporting scenarios of ownership transfer and the extension of DPP information for complex products.

Drawing upon the specific requirements outlined in the European Regulation for Ecodesign of Sustainable Products (ESPR), a comprehensive analysis of each functional requirement of the Digital Product Passport is conducted. In this analysis, various system elements that could support the DPP are proposed, and implementation alternatives are explored, considering the technological ambiguity present in these European regulations.

% fmunyoz: The elements that compose a paper are sections, instead of chapters. Besides, in the main file we assigned a label for each section.

%Chapter five 
Section \ref{sec:proposal} clusters the diverse design options into two viable proposals for the DPP: one based on DID and the other on verifiable credentials. After presenting both alternatives, a detailed comparison is conducted to determine which one is more suitable for implementation. The selection of the most appropriate proposal will depend on various factors, including technical, economic, legal, and acceptance aspects from different stakeholders. The final decision must carefully consider the advantages and disadvantages of each proposal in relation to the specific requirements of the Digital Product Passport and the overall goals of the circular economy in the European Union.

Ultimately, this article offers a comprehensive and well-founded perspective on the potential of decentralised identity for implementing the Digital Product Passport, laying the groundwork for future research and discussions on applying innovative technologies towards a more sustainable and circular economy.

\bibliographystyle{plain}
\bibliography{main}

\end{document}